\documentclass[journal,10pt]{IEEEtran}
\usepackage[T1]{fontenc}
\usepackage{cite}
\usepackage{amsmath,amssymb}
\usepackage{graphicx}
\usepackage{booktabs,array,tabularx,multirow}
\usepackage[table]{xcolor}
\usepackage{xurl}
\usepackage{flushend}
\usepackage[hidelinks]{hyperref}

\hypersetup{pdftitle={RealComm: Benchmarking and Adapting Audio Deepfake Detection over Real Communication Channels}}
\graphicspath{{assets/figures/}}

\newcommand{\eer}{\operatorname{EER}}
\begin{document}
\title{\fontsize{22}{26}\selectfont RealComm: Benchmarking and Adapting Audio\\Deepfake Detection over Real Communication Channels}
\author{Li~Wang$^{*}$, Jindong~Wang$^{*}$, Wan~Lin$^{*}$, Kunyu~Feng$^{*}$, Yufan~Lin, Lei~Wang,\\
Rizhao~Cai, Ce~Fang, Jinzhe~Xue, Jie~Shi, Haizhou~Li, and Zhizheng~Wu%
\thanks{$^{*}$Li Wang, Jindong Wang, Wan Lin, and Kunyu Feng contributed equally to this work.}%
\thanks{Li Wang, Jindong Wang, Wan Lin, Kunyu Feng, Yufan Lin, Haizhou Li, and Zhizheng Wu are with The Chinese University of Hong Kong, Shenzhen, China (e-mail: liwang1@link.cuhk.edu.cn; wuzhizheng@cuhk.edu.cn).}%
\thanks{Zhizheng Wu is also with Amphion Technology Co., Ltd.}%
\thanks{Lei Wang, Rizhao Cai, Ce Fang, Jinzhe Xue, and Jie Shi are with Huawei Technologies Co., Ltd.}}
\maketitle

\begin{abstract}
Acoustic capture, device processing, and transmission coding in real calls jointly alter speech signals and may affect the reliability of audio deepfake detection. To systematically evaluate these effects, we introduce RealComm, a dataset pairing digital speech with recordings of the same utterances transmitted through real mobile calls. The acquisition design covers acoustic and wired injection, handset and speakerphone operation, and caller-side denoising, with bona fide and synthetic speech recorded under matched conditions. The dataset includes RealCommBench for evaluation and RealCommTrain for adaptation. We further develop RealComm-Aug, a staged waveform augmentation pipeline for simulating acquisition and transmission effects, and investigate its combination with real-call training. Experiments with existing detectors reveal substantial degradation after call transmission, even for models with strong digital-domain performance; acoustic injection is particularly challenging. Paired quality and spectral analyses show that quality degradation is associated with poorer detection across recording conditions for the evaluated detectors, while high-frequency attenuation alone does not explain detection difficulty. Simulated augmentation and real-call training provide complementary benefits, with their combination achieving the lowest call-domain error for each of three evaluated detectors. For Teffic-Audio, the equal error rate decreases from 37.94\% to 22.37\%. The demo and benchmark access information are provided through the project repository.
\end{abstract}

\section{Introduction}
\IEEEPARstart{A}{udio} deepfakes are speech recordings synthesized or manipulated using machine learning, including text-to-speech synthesis and voice conversion~\cite{survey}. Recent systems can produce natural speech resembling a target speaker from text and a short reference recording~\cite{seedtts,cosyvoice2,f5tts}. These capabilities can also be misused for speaker impersonation and voice-based fraud. Audio deepfake detection addresses this risk by distinguishing bona fide speech from artificially generated or manipulated speech, as reviewed by Yi et al.~\cite{survey}.

\textbf{Real calls expose detectors to a more complex signal path than direct evaluation of digital audio.} A waveform may pass through loudspeaker playback and microphone capture before device processing, denoising, and transmission coding modify it further. These operations can alter spectral and temporal cues used by detectors and reduce their reliability~\cite{channel}. Evaluating the digital source alone therefore leaves an important question unanswered: how well does the detector perform on the speech actually received in a call?

\textbf{We introduce RealComm to study this gap using paired digital and real-call recordings.} The acquisition design covers acoustic and wired injection, handset and speakerphone operation, multiple caller--receiver device combinations, and user-controlled smartphone denoising. Bona fide and synthetic speech traverse the same recording conditions. Pairing each call recording with its digital source holds content, speaker, and generation system fixed when comparing the two domains. RealCommBench supports evaluation across these conditions, while RealCommTrain provides recordings for adaptation; all versions of a source remain within one partition.\footnote{An earlier version of this work was accepted at IEEE SLT~\cite{realcomm_slt}. Project repository: \url{https://github.com/AmphionTeam/RealComm}.}

\textbf{Existing detectors show substantial degradation after real-call transmission.} Condition-level evaluation identifies acoustic injection as particularly challenging, while the effect of denoising varies with the detector and input configuration. We further analyze paired speech-quality and spectral changes to understand what accompanies these failures. Quality deterioration tracks detection degradation for the evaluated detectors, whereas high-frequency attenuation alone does not explain differences in condition difficulty. These findings motivate adaptation that accounts for multiple stages of the communication path.

\textbf{We address this mismatch with RealComm-Aug and real-call adaptation.} RealComm-Aug organizes acoustic, device, coding, and transmission perturbations by their position in a plausible signal path. Intermediate outputs retain common noise, reverberation, and file-compression augmentation without requiring every example to undergo a simulated call. The pipeline operates on input waveforms and can be used with existing detectors while retaining their model-internal augmentation. We evaluate simulated augmentation, training with RealCommTrain, and their combination from the same initial detector checkpoints. Their combination achieves the lowest pooled call-domain error for each of the three evaluated detectors.

Our contributions are as follows:
\begin{itemize}
\item A paired real-call benchmark and adaptation set covering injection paths, device combinations, handset/speakerphone operation, and denoising settings.
\item An evaluation of existing detectors relating detection degradation to acquisition conditions, speech quality, and spectral changes.
\item A staged augmentation pipeline and a three-model comparison of simulated and real-call adaptation.
\end{itemize}

\section{Related Work}
\subsection{Benchmarks and Generalization}
Audio deepfake benchmarks provide complementary coverage of attacks and acquisition settings. ASVspoof 2019 distinguishes logical-access synthesis and conversion from physical-access replay~\cite{asvspoof19}. ASVspoof 2021 adds transmission and compression variation~\cite{asvspoof21}, while ASVspoof 5 broadens speaker, recording, and attack diversity through crowdsourced resources~\cite{asvspoof5}. Its evaluation further examines adversarial attacks and neural compression~\cite{asvspoof5eval}.

Other datasets broaden source material and generation mechanisms. WaveFake covers neural waveform generators~\cite{wavefake}, MLAAD emphasizes multilingual synthesis~\cite{mlaad}, and Codecfake addresses speech associated with neural audio codecs~\cite{codecfake}. In-the-wild evaluation highlights cross-dataset generalization challenges~\cite{inthewild}; SpoofCeleb incorporates diverse real-world source conditions into generation and detection evaluation~\cite{spoofceleb}. Generalization is also a central theme of the survey by Yi et al.~\cite{survey}. \textbf{RealComm complements this coverage with paired source and real-call recordings.} Holding the source utterance fixed supports comparisons across acquisition conditions.

\subsection{Channel Effects and Real Communication}
Zhang et al. examine channel mismatch and study augmentation, multi-task learning, and adversarial learning as remedies~\cite{channel}. Broader corruption studies analyze detector responses to noise, signal modification, and compression~\cite{robustness}. These controlled experiments characterize sensitivity to specific transformations.

Communication-focused studies examine further aspects of transmission. ADD-C evaluates codec compression and simulated packet loss, together with communication-oriented augmentation~\cite{addc}. RTCFake studies transmitted speech across real-time communication platforms~\cite{rtcfake}. Tsutsumi et al. evaluate phone calls, Telegram calls, and interviews, and demonstrate the value of mixing original and degraded speech for adaptation~\cite{degraded}. Our earlier RealComm benchmark also studies detection over real communication channels~\cite{realcomm_slt}. \textbf{The present acquisition design combines input paths, device routes, call modes, and denoising states.} Paired recordings connect these conditions to detection, speech-quality measurements, and adaptation.

\subsection{Detector Representations}
End-to-end detectors such as RawNet2 learn directly from waveforms~\cite{rawnet2}; AASIST combines spectral and temporal graph attention~\cite{aasist}. Another family builds on pretrained speech representations. Wav2vec 2.0 and HuBERT establish self-supervised representation learning~\cite{wav2vec2,hubert}, with XLS-R extending its multilingual scale~\cite{xlsr} and WavLM covering a wider range of speech tasks~\cite{wavlm}. Whisper learns through large-scale weak supervision~\cite{whisper}, while the Seamless work provides the W2v-BERT 2.0 encoder used by Teffic-Audio~\cite{seamless}.

Pretrained encoders support authenticity assessment through a detection head and task-specific training. Tak et al. combine a fine-tuned wav2vec 2.0 front end with augmentation for anti-spoofing~\cite{sslaug}, and XLSR-SLS selects useful self-supervised layers for detection~\cite{sls}. Our comparison includes detectors from these families and the released DF Arena model~\cite{dfarena}. \textbf{We evaluate complete detection systems under one paired protocol.} The results reflect their architectures together with their pretraining and detection-training recipes.

\subsection{Augmentation, Adaptation, and Quality Diagnostics}
RawBoost introduces convolutive and additive waveform perturbations motivated in part by telephony~\cite{rawboost}. Cohen et al. study compression and channel augmentation for voice anti-spoofing~\cite{cohenaug}. Neural audio compression, including EnCodec~\cite{encodec}, provides additional training transformations alongside conventional coding perturbations. Together with real-channel adaptation studies~\cite{degraded}, these methods motivate comparing simulated variation, actual call recordings, and their combination.

\textbf{RealComm connects augmentation and quality diagnostics through paired evaluation.} RealComm-Aug organizes established transformations along plausible acquisition and communication paths and is evaluated both alone and with real-call training. Seed-TTS evaluation measures content accuracy and prompt-referenced speaker similarity~\cite{seedtts,seedttseval}, while DNSMOS P.835 predicts perceptual speech quality~\cite{dnsmos}. These diagnostics describe complementary aspects of speech quality; we examine how their changes relate to detector degradation.

\section{The RealComm Corpus}
\label{sec:data}
\subsection{Call Acquisition and Conditions}
\begin{figure*}[t]
\centering
\includegraphics[width=\textwidth,trim=0bp 2bp 0bp 2bp,clip]{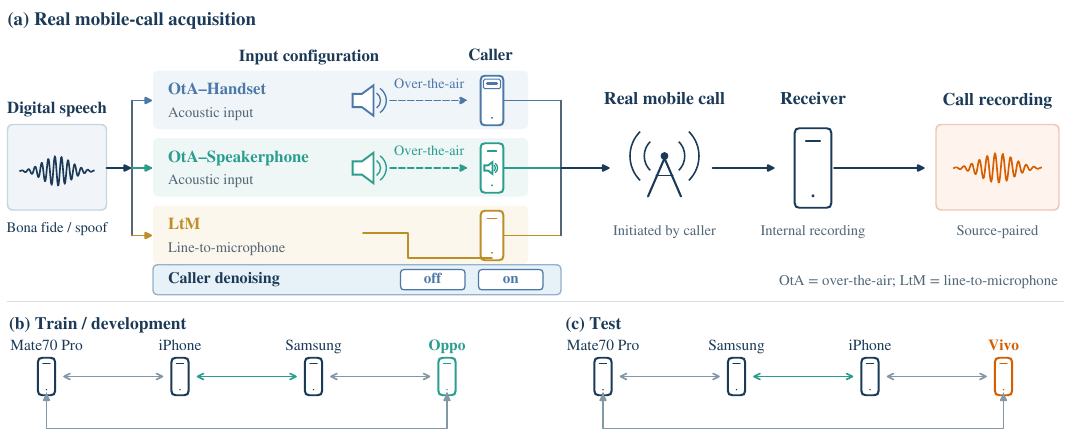}
\caption{RealComm acquisition and device partitions. The upper panel shows acoustic and wired injection into the caller, followed by a real mobile-call link and receiver-side recording. The lower panels show the bidirectional training/development and test rings; the return edge closes each ring. The Samsung--iPhone pair is shared; Oppo appears only in training/development and Vivo only in testing. Call modes and denoising settings refer to the caller.}
\label{fig:dataset}
\end{figure*}

Figure~\ref{fig:dataset} summarizes the acquisition chain. Digital source speech is injected into a caller device, which initiates the call. The receiver records the received signal using the calling application's recording function. This setup produces a digital reference and multiple call recordings for each source utterance.

\subsubsection{Input configuration}
Over-the-air (OtA) injection plays a source waveform through an external loudspeaker, allowing it to propagate through air into the caller's microphone. We record both handset and speakerphone operation of the caller. Line-to-microphone (LtM) injection connects the waveform to the microphone input through a cable, bypassing the additional air-propagation segment. The resulting configurations are OtA--Handset, OtA--Speakerphone, and LtM. ``Handset'' denotes the call operating mode; it is not an additional source or a different test partition.

For OtA playback, a Huawei Sound X loudspeaker and the caller phone were positioned at the same height, approximately 1~m above the floor. The nominal loudspeaker volume and separation were approximately 60\% and 30~cm, respectively; playback level and spacing were adjusted as needed to obtain clear recordings. LtM delivered the signal through the phone's wired-headset microphone input.

\subsubsection{Directed device routes}
Each route is directed from the caller supplying the source speech to the receiver recording it; reversing these roles defines a different route. The training/development ring is Mate70 Pro--iPhone--Samsung--Oppo--Mate70 Pro. The test ring is Mate70 Pro--Samsung--iPhone--Vivo--Mate70 Pro. Adjacent devices communicate in both directions, giving eight directed routes per ring. Compared with all twelve directed pairs among four devices, the ring reduces acquisition requirements while retaining two neighbors per device in both roles. Its partial overlap permits comparisons between a shared route, an unseen pairing of covered devices, and an unseen device in either role.

\subsubsection{User-controlled denoising}
Smartphone denoising can alter signal cues used by deepfake detectors while suppressing background noise, motivating paired evaluation of its on and off states. Eight routes and three input configurations give 24 base conditions. For configurations supporting a user-controlled denoising switch, we additionally record the on state. This adds 20 training-side and 18 test-side conditions, for totals of 44 and 42. We retain three distinct labels: \emph{off}, \emph{on}, and \emph{unavailable}. The last means that no controllable switch exists in that configuration; it is not treated as off. Conversely, off means only that the exposed function is disabled, not that all internal speech processing has been removed. Table~\ref{tab:devices} lists the device models, recording-time software versions, and denoising controls available in the recorded configurations.

\begin{table}[t]
\centering
\caption{Devices and caller-side denoising controls.}
\label{tab:devices}
\footnotesize
\setlength{\tabcolsep}{2pt}
\begin{tabularx}{\columnwidth}{@{}p{.47\columnwidth} X c c c@{}}
\toprule
OS & Denoising & H & S & L\\
\midrule
\rowcolor{black!8}[0pt][0pt]
\multicolumn{5}{@{}c@{}}{\strut\textbf{Huawei Mate70 Pro (Mate70 Pro)}}\\
HarmonyOS 6.1.0.135 SP8 & Call noise reduction & Yes & Yes & Yes\\[2pt]
\rowcolor{black!8}[0pt][0pt]
\multicolumn{5}{@{}c@{}}{\strut\textbf{iPhone 15 Pro (iPhone)}}\\
iOS 26.6.1 & Voice Isolation & Yes & Yes & Yes\\[2pt]
\rowcolor{black!8}[0pt][0pt]
\multicolumn{5}{@{}c@{}}{\strut\textbf{Samsung Galaxy S25+ (Samsung)}}\\
Android 16 & Voice focus & Yes & Yes & NA\\[2pt]
\rowcolor{black!8}[0pt][0pt]
\multicolumn{5}{@{}c@{}}{\strut\textbf{Oppo Find X9 (Oppo)}}\\
ColorOS 16.0.1 & Clear Voice & Yes & Yes & NA\\[2pt]
\rowcolor{black!8}[0pt][0pt]
\multicolumn{5}{@{}c@{}}{\strut\textbf{Vivo X300 Pro (Vivo)}}\\
OriginOS 6 & Voice Isolation & NA & Yes & NA\\[2pt]
\bottomrule
\end{tabularx}
\par\vspace{2pt}
\parbox{\columnwidth}{\scriptsize OS versions are those used for recording. H: OtA--Handset; S: OtA--Speakerphone; L: LtM. Yes: paired off/on recordings. NA: no user-controlled switch.}
\end{table}

\subsection{Source Speech and Partitioning}
The digital pool contains 2,800 utterances, balanced between bona fide and synthetic speech and between English and Mandarin. Bona fide speech, target text, and reference prompts originate from Seed-TTS Eval~\cite{seedtts,seedttseval}. Synthetic speech is generated by CosyVoice2~\cite{cosyvoice2}, F5-TTS~\cite{f5tts}, FlexiVoice~\cite{flexivoice}, IndexTTS2~\cite{indextts2}, MaskGCT~\cite{maskgct}, Vevo2~\cite{vevo2}, and Minimax~\cite{minimax}, with 200 utterances per system. Source selection uses content-accuracy and audio-quality screening; detector scores are not used to select samples. Consequently, the source pool is a quality-screened cohort rather than an unrestricted sample of all generation outcomes.

We group related content and reference-audio identities before partitioning. An initial 8:2 allocation reserves 560 digital sources for testing and 2,240 for the training side. A second 8:2 allocation divides the latter into 1,792 train and 448 development sources. All digital and call versions of a source inherit the same partition. This prevents a recorded version of an utterance from crossing into a different split from its digital counterpart.

The acquisition subset contains 224 train and 56 development sources, each recorded under 44 conditions. All 560 test sources are recorded under 42 conditions. Table~\ref{tab:splits} distinguishes independent source counts from recording counts. RealCommTrain comprises the train and development recordings, while RealCommBench comprises the test recordings. The smaller source set on the training side controls the cost of repeated real acquisition; the complete digital training pool remains available for other protocols.

\begin{table}[t]
\centering
\caption{Source-level partitions and real-call recording counts.}
\label{tab:splits}
\begin{tabular}{lrrrr}
\toprule
Split & Digital & Recorded & Conditions & Call\\
 & sources & sources & per source & recordings\\
\midrule
Train & 1,792 & 224 & 44 & 9,856\\
Dev & 448 & 56 & 44 & 2,464\\
Test & 560 & 560 & 42 & 23,520\\
\bottomrule
\end{tabular}
\end{table}

The train and test sets cover the same seven generation systems. The intended generalization axes are held-out source content and communication paths, not unseen generators. Multiple recordings of a source expand condition coverage but do not add independent content. The balanced source and acquisition design also ensures that neither authenticity class is uniquely associated with a recording condition.

\subsection{One Test Set, Several Analysis Dimensions}
All 23,520 call recordings form a single test set. We report its pooled EER as the principal call-domain result. Input configuration, denoising state, and directed route are annotations for analyzing that same set, rather than alternative test protocols. OtA--Handset, OtA--Speakerphone, and LtM contain 7,840, 8,960, and 6,720 recordings, respectively. Their different sizes reflect the availability of denoising switches.

For adaptation analysis, we summarize routes in four groups relative to the RealCommTrain ring: Samsung$\leftrightarrow$iPhone (seen route), Mate70 Pro$\leftrightarrow$Samsung (unseen pairing), and Vivo as caller or receiver. ``Seen'' refers to the real-training acquisition design; it does not mean that the zero-shot models have encountered those calls. Group differences may also reflect the mixture of input and denoising conditions, so they are not interpreted as isolated hardware effects.

\section{Experimental Protocol}
\label{sec:protocol}
\subsection{Models and Waveform Input}
The main zero-shot analysis uses five detectors: WavLM-L (Teffic), Whisper (Teffic), AASIST (Teffic), Teffic-Audio, and DF Arena 500M V1. We shorten the first three names to WavLM-L, Whisper, and AASIST, and the last to DF Arena. Teffic-Audio uses a W2v-BERT 2.0 encoder with an utterance-level pooling and classification head. These are existing trained detectors, not speech encoders used without a detection head. The existing-weight condition is denoted F and performs no RealComm parameter updates. Subsequent adaptation experiments use WavLM-L, AASIST, and Teffic-Audio. An initial evaluation also included XLSR-SLS, whose digital/call EERs were 38.57\%/50.31\%; its weak digital baseline limits interpretation of additional channel degradation, so it is not included in the main figures and comparisons.

Each model evaluates all 24,080 files: 560 digital test sources and 23,520 paired call recordings. Audio is converted to mono at 16~kHz. WavLM-L, Whisper, and Teffic-Audio read at most the first 64,600 samples; shorter utterances retain their variable length. AASIST and DF Arena first read the leading 64,000 samples and then cyclically pad to 64,600 samples as required by their interfaces. Each file receives one forward pass, with no sliding-window averaging. These model-specific rules are used consistently within each digital/call comparison. Cross-model differences therefore include both the trained system and its input interface.

We save one scalar score and one embedding from the same inference pass per file. Scores are oriented so that larger values indicate synthetic speech. Embeddings retain the recorded extraction layer and pooling convention without additional normalization. The present main analysis uses detection scores; embeddings support subsequent within-model diagnostics and are not used to select test samples.

\subsection{Metrics and Matched Comparisons}
We report EER in percent. Let $s_m(x)$ be the spoof-oriented score of model $m$ on waveform $x$. For digital sources $\mathcal{D}$ and call recordings $\mathcal{C}$, the domain gap is
\begin{equation}
\Delta_m = \eer_m(\mathcal{C})-\eer_m(\mathcal{D}),
\end{equation}
expressed in percentage points. The pooled call EER is computed directly from all scores in $\mathcal{C}$, not by averaging subgroup EERs. A per-generator EER uses that generator's spoof samples and the same-domain bona fide reference set.

Input-path comparisons are supplemented by matched analyses using the same sources, routes, and shared denoising labels. Denoising off/on comparisons use 10,080 pairs with identical source, directed route, and input configuration. The 3,360 unavailable recordings contribute to the full test result but not to switch comparisons. When estimating uncertainty in domain differences, we use 2,000 paired bootstrap replicates over 148 source-association clusters, retaining all digital and call versions in each sampled cluster. These intervals characterize source uncertainty under the fixed acquisition conditions, not variability across new devices or independent recording campaigns.

All adaptation results are single-run checkpoint results. We preserve full-precision metrics for differences and round displayed EERs to two decimals. Thus a difference calculated from full-precision scores can differ by 0.01 points from subtracting two displayed values.

\section{Zero-Shot Detection Results}
\label{sec:zero}
\begin{figure*}[t]
\centering
\includegraphics[width=\textwidth]{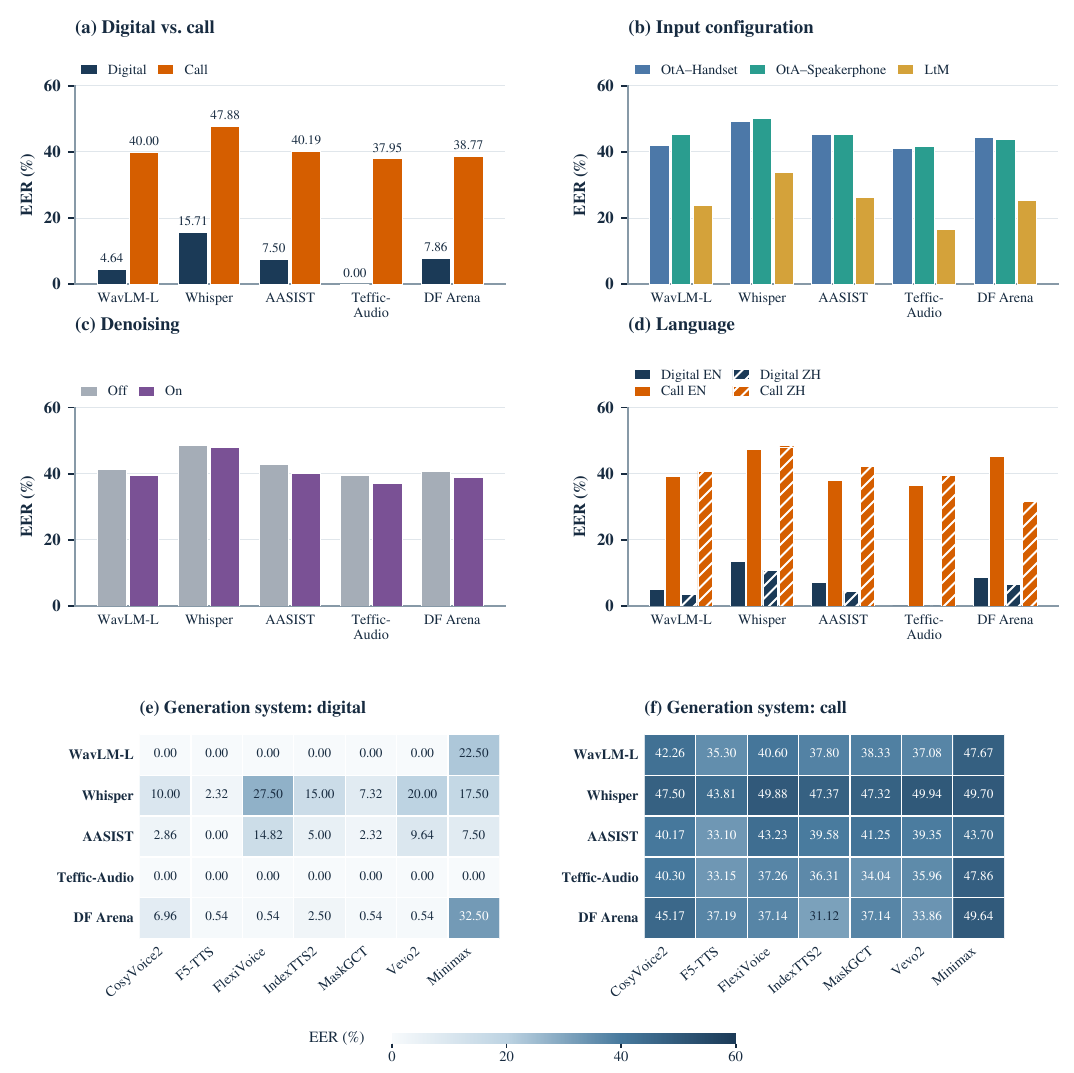}
\caption{Zero-shot EER (\%, lower is better): (a) paired digital and call domains; (b) input configuration; (c) matched denoising off/on; (d) English (EN) and Mandarin (ZH) in both domains; (e,f) generation systems in digital and call speech. Bar panels and generation-system heatmaps share a 0--60\% range. Each generator is compared with the same-domain bona fide set. All panels use the same detectors.}
\label{fig:zero}
\end{figure*}

\subsection{A Large Digital-to-Call Gap}
\textbf{Real-call transmission substantially degrades the evaluated detectors.} In Fig.~\ref{fig:zero}(a), every model loses more than 30 EER points despite digital EERs of 0--15.71\%. WavLM-L rises from 4.64\% to 40.00\%, AASIST from 7.50\% to 40.19\%, and DF Arena from 7.86\% to 38.77\%. Teffic-Audio reaches 37.95\% from 0.00\%. Because the domains contain the same source utterances, this gap cannot be explained simply by a different selection of generated speech.

\subsection{Acoustic and Wired Injection}
\textbf{Acoustic injection is consistently more difficult than wired injection.} Every detector has lower EER on LtM than on either OtA configuration in Fig.~\ref{fig:zero}(b). WavLM-L reaches 23.72\% on LtM, compared with 41.96\% on handset and 45.33\% on speakerphone; Teffic-Audio reaches 16.46\%, 41.10\%, and 41.70\%, respectively. The LtM advantage persists after matching sources, routes, and denoising labels, linking the additional playback-and-capture segment to greater detection difficulty. The recordings do not isolate the contributions of the loudspeaker, room, microphone, and internal processing.

\textbf{Neither handset nor speakerphone is consistently easier to detect.} Their EER ordering varies across models. The robust distinction in this comparison is the gap between acoustic and wired injection, rather than a general advantage of either call mode.

\subsection{Denoising, Language, and Generation Systems}
\textbf{Denoising lowers pooled EER across the evaluated detectors.} WavLM-L, Whisper, AASIST, Teffic-Audio, and DF Arena all improve on the matched off/on subsets in Fig.~\ref{fig:zero}(c). For example, AASIST falls from 42.70\% to 40.18\%. This shared pooled direction does not imply improvement for every route and input configuration; the individual paired conditions include both gains and regressions.

\textbf{The call-domain degradation occurs in both languages.} Figure~\ref{fig:zero}(d) shows higher EER after transmission in English and Mandarin for every detector. Language ordering still depends on the model: AASIST has lower call EER in English, whereas DF Arena has lower EER in Mandarin. Each cohort contains 280 digital sources and 11,760 call recordings; differences in speakers and content prevent attributing the ordering to language alone.

\textbf{The degradation extends across all seven evaluated generation systems.} Each evaluated detector deteriorates for every generator in Fig.~\ref{fig:zero}(e,f). The effect is therefore not confined to one synthesizer. These comparisons share the same-domain bona fide reference set and establish coverage within the sampled generators; they do not evaluate unseen-generator generalization.

\section{Speech Quality, Spectral Changes, and Detection}
\label{sec:quality}
\subsection{Paired Speech-Quality Evaluation}
Content accuracy and speaker similarity follow Seed-TTS Eval~\cite{seedttseval}. English WER is computed with Whisper-large-v3 and Mandarin CER with Paraformer-zh~\cite{paraformer}. Both the digital and call waveform are evaluated against the same target text. Reported WER/CER values are means of per-utterance error rates, rather than error counts pooled over all words or characters.

Speaker similarity (SIM) uses the official WavLM-large speaker-verification model. For a reference prompt $p_i$ and source $x_i$, we compare $\cos(e(p_i),e(x_i))$ with $\cos(e(p_i),e(x_{i,c}))$, where $x_{i,c}$ is the call recording under condition $c$. The reference prompt is held fixed across domains. SIM is reported separately for bona fide and synthetic speech; it measures similarity to that prompt, not waveform fidelity to the digital source. The evaluators are separate models from the authenticity detectors, although they can share a backbone family.

Overall quality is measured by DNSMOS P.835 OVRL~\cite{dnsmos}. This is a non-intrusive predictor of overall speech quality, evaluated independently on each waveform. Digital speech supplies the paired baseline for its change; it is not an input reference to DNSMOS itself. We use these automatic metrics to characterize the recordings, not as substitutes for subjective listening or security performance.

\begin{figure*}[t]
\centering
\includegraphics[width=\textwidth,trim=0bp 10.5bp 0bp 16.5bp,clip]{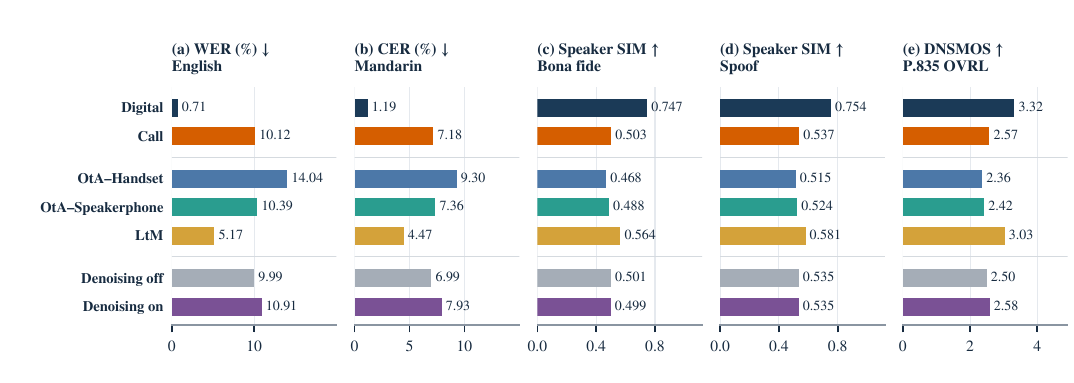}
\caption{Mean quality metrics in the digital and call domains, by input configuration, and for matched denoising off/on subsets. Each metric retains its own scale. WER and CER are averaged over utterances in the corresponding language. SIM is prompt-referenced and separated by authenticity class. DNSMOS denotes P.835 OVRL. Unavailable denoising cases remain in the pooled call result but are omitted from the switch panels.}
\label{fig:quality}
\end{figure*}

\textbf{Call transmission reduces content accuracy, speaker similarity, and predicted quality.} Mean English WER rises from 0.71\% to 10.12\%, and Mandarin CER from 1.19\% to 7.18\%. Bona fide SIM falls from 0.747 to 0.503, synthetic SIM from 0.754 to 0.537, and DNSMOS from 3.32 to 2.57.

\textbf{LtM preserves higher measured quality than either OtA configuration.} It has the lowest recognition error and the highest SIM and DNSMOS in Fig.~\ref{fig:quality}. Its DNSMOS is 3.03, compared with 2.36 and 2.42 for handset and speakerphone. This input-path ordering agrees with the detection results in Section~\ref{sec:zero}.

\textbf{The quality effect of denoising depends on the input configuration.} On four common routes, enabling denoising raises speakerphone WER and CER by 3.96 and 3.48 points, but lowers the corresponding LtM means by 2.72 and 1.38 points. Both SIM measures likewise decrease on speakerphone and increase on LtM. Thus, pooled quality means can hide opposite changes across configurations; these observations describe the recorded routes rather than a universal switch recommendation.

\subsection{Frequency-Dependent Changes}
\begin{figure}[t]
\centering
\includegraphics[width=\columnwidth,trim=0bp 3bp 0bp 2bp,clip]{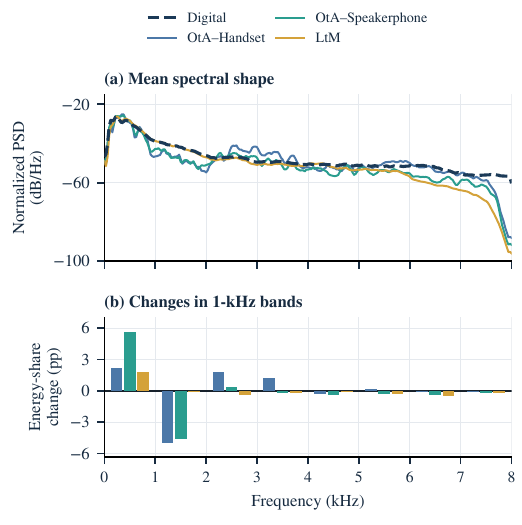}
\caption{Paired spectral changes. Top: normalized mean power spectral density (PSD). Bottom: changes in the energy share of successive 1-kHz bands, in percentage points relative to digital speech. The panels share a 0--8~kHz axis. The digital curve is dark blue and dashed.}
\label{fig:spectrum}
\end{figure}

We estimate the PSD of each utterance using a common analysis procedure at 16~kHz and normalize it by its total power. For each input configuration, call PSDs are first averaged within source, then equally across the 560 sources. Averaging occurs in the linear domain before conversion to decibels. This procedure describes changes in spectral shape while preventing a source with more recordings from receiving greater weight.

\textbf{Real calls alter spectral shape, with pronounced attenuation near the upper analysis limit.} All three configurations show a sharp decline around 7.5--8~kHz in Fig.~\ref{fig:spectrum}. The 7--8~kHz energy share falls from 0.248\% in digital speech to 0.123\%, 0.058\%, and 0.015\% for handset, speakerphone, and LtM. These normalized shares describe spectral shape, not absolute gain or a specific negotiated codec cutoff.

\textbf{Acoustic injection also changes the speech midband.} The 1--2~kHz energy share falls from 6.81\% to 1.84\% and 2.25\% for handset and speakerphone, while LtM remains at 6.79\%. This band overlaps part of the vowel-formant region associated with phonetic distinctions~\cite{hillenbrand1995}. Its relative reduction in OtA suggests that the additional playback-and-capture path may reshape the speech spectral envelope and alter the prominence of these cues, rather than only removing the high-frequency tail. This is a plausible interpretation of the observed spectral redistribution, not a direct measurement of formant loss.

\textbf{High-frequency preservation alone does not explain the input-path detection ranking.} LtM has the smallest 7--8~kHz energy share while yielding the lowest EER for every detector. The high-frequency tail therefore does not provide a sufficient explanation for why OtA is harder to detect.

\subsection{Condition-Level Association with Detection Degradation}
Each of the 42 conditions contains the same 560 test sources. For a quality measure $q$, we compute a condition-level mean deterioration
\begin{equation}
d_{q,c}=\frac{1}{N_q}\sum_{i\in\mathcal{I}_q}\epsilon_q\big[q(x_{i,c})-q(x_i)\big],
\label{eq:qualitychange}
\end{equation}
where $\epsilon_q=+1$ for WER/CER and $-1$ for SIM/DNSMOS. $\mathcal{I}_q$ contains the applicable sources: the corresponding language for WER/CER and all sources for the other metrics. We correlate $d_{q,c}$ across conditions with
\begin{equation}
\Delta_{m,c}=\eer_m(\mathcal{C}_c)-\eer_m(\mathcal{D})
\end{equation}
using Spearman's rank coefficient. WER/CER use language-matched EERs. A positive coefficient means that conditions with larger quality deterioration tend to exhibit greater detection deterioration. For spectral analysis, we substitute the decrease in normalized 4--8~kHz energy share; this is a spectral change, not intrinsically a quality score.

\begin{figure}[t]
\centering
\includegraphics[width=\columnwidth,trim=0bp 18bp 0bp 0bp,clip]{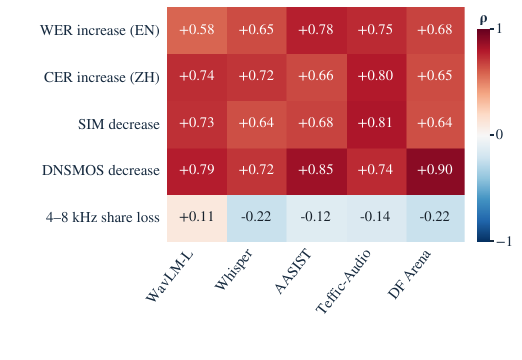}
\caption{Spearman association across 42 recording conditions between paired quality or spectral changes and EER increases. WER and CER use language-matched EERs; SIM pools both authenticity classes equally. The high-frequency row describes a reduction in relative energy, not necessarily worse perceived quality.}
\label{fig:association}
\end{figure}

\textbf{Quality deterioration consistently tracks detection degradation.} Across the evaluated detectors, WER, CER, SIM, and DNSMOS deterioration have positive condition-level correlations with EER increases in Fig.~\ref{fig:association}. Across models, the coefficients range from 0.58 to 0.78 for WER, 0.65 to 0.80 for CER, 0.64 to 0.81 for SIM, and 0.72 to 0.90 for DNSMOS. This is a consistent descriptive relationship across the evaluated quality measures and detectors.

\textbf{The quality association is not solely a difference between acoustic and wired injection.} Within LtM, DNSMOS deterioration remains positively associated with EER increases for the same detectors, with coefficients of 0.54--0.87. The relationship also varies within a shared input path, although its strength depends on the detector and configuration.

\textbf{High-frequency loss only weakly tracks detection degradation.} Across the evaluated detectors, the correlations for the 4--8~kHz energy-share decrease range from $-0.22$ to $0.11$ and differ in sign. The contrast with the quality measures shows that visible spectral attenuation alone does not reliably rank condition difficulty.

\textbf{These correlations describe conditions, not causal effects or individual decisions.} Conditions share sources and devices, and quality and detection can both respond to the processing chain. Quality metrics also use complete utterances, while detection follows the prefix protocol of Section~\ref{sec:protocol}. The observed agreement therefore supports quality-aware analysis of the benchmark, rather than using an automatic quality score as a replacement for detection evaluation.

\section{Simulated Augmentation and Real-Call Adaptation}
\label{sec:method}
\subsection{RealComm-Aug}
\begin{figure*}[t]
\centering
\includegraphics[width=\textwidth]{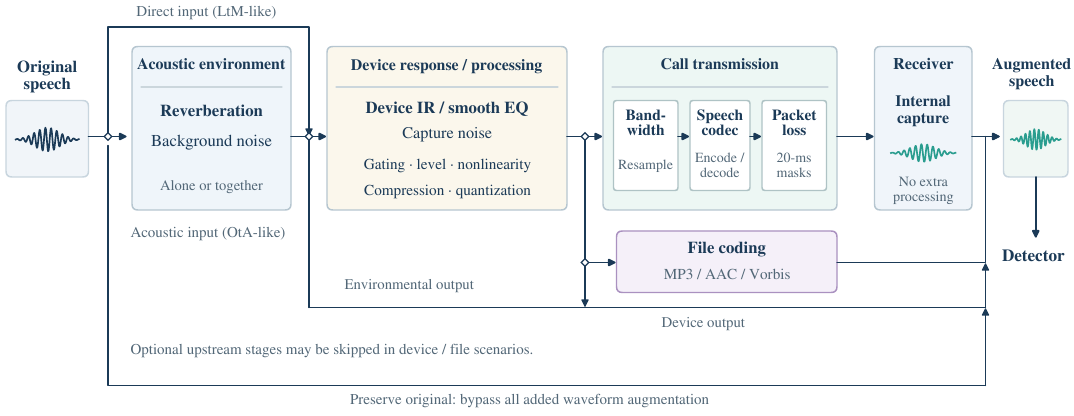}
\caption{RealComm-Aug processing order and output scenarios. One scenario determines the stages and output exit; individual operations are sampled within that scenario. The upper bypass represents direct, LtM-like input, while the acoustic path passes through the environmental stage. Environmental, device, and file outputs provide non-call augmentation. Upstream stages may be skipped in device/file scenarios. Call output is captured internally without receiver-side replay. IR: impulse response; EQ: equalization.}
\label{fig:aug}
\end{figure*}

RealComm-Aug applies waveform transformations in the order of acquisition and communication. For each input, it first selects one processing scenario using predefined probabilities, then samples the operations and parameters within that scenario. Bona fide and synthetic speech share the same sampling distribution, and the original authenticity label is preserved. Figure~\ref{fig:aug} groups the operations into acoustic environment, device response and processing, file coding, and call transmission. Intermediate outputs retain conventional noise, reverberation, device, and file augmentation without requiring every example to undergo a simulated call.

\subsubsection{Acoustic environment}
Room impulse responses from RIRS\_NOISES and Graphi07 simulate reverberation, while MUSAN~\cite{musan} and Gaussian noise supply background disturbances. Reverberation and background noise can operate independently or together. Environmental background sound is introduced before downstream device and communication processing, so it is transformed together with the speech. The direct, LtM-like call path bypasses this additional air-propagation stage.

\subsubsection{Device response and processing}
Device impulse responses from micirp, MADIR, and POLIPHONE, or a smooth equalization curve, perturb the frequency response. Capture noise is distinct from environmental background sound. Low-amplitude gating, soft clipping or convolutive nonlinear perturbation, level control, dynamic compression, and linear bit-depth quantization provide amplitude and nonlinear variability. On the direct call path, optional equalization is used without an additional acoustic microphone response. These operations approximate families of device effects; they do not reproduce a particular RealComm phone model.

\subsubsection{File and communication coding}
MP3, AAC, and Vorbis form the file-coding branch. The call branch combines frequency filtering, round-trip resampling, and candidates drawn from GSM, Opus, Speex, G.711, and G.723.1. File and communication coding are sampled as different output scenarios rather than sequentially imposed on every example. The codec candidates create training distortions and are not claims about the codecs negotiated in the real recordings.

\subsubsection{Packet-loss simulation and receiver output}
Independent or burst masks operate on 20-ms decoded waveform frames, replacing masked segments with silence or a repeated preceding frame. This approximates audible consequences of packet loss; it does not discard actual encoded packets or reproduce a complete network stack. The received waveform is then passed directly to the detector. There is no receiver-side loudspeaker replay, room convolution, or added environmental noise.

\subsection{Training Conditions}
\begin{table*}[t]
\centering
\caption{Adaptation settings. ``Original digital data'' denotes the detector's existing training mixture, not the digital counterparts of RealCommTrain. Every adapted model starts from its corresponding F checkpoint.}
\label{tab:settings}
\begin{tabularx}{\textwidth}{l p{.28\textwidth} X}
\toprule
Setting & Training data & Data-side waveform processing\\
\midrule
F & No parameter updates & Evaluation preprocessing only\\
A & Original digital data & RealComm-Aug\\
T & Original digital data + RealCommTrain/train & Original augmentation on digital examples; no additional data-side augmentation on call recordings\\
A+T & Same data as T & RealComm-Aug on digital examples; no additional data-side augmentation on call recordings\\
\bottomrule
\end{tabularx}
\end{table*}

Table~\ref{tab:settings} defines the comparisons. A adapts the detector using simulated perturbations of its original digital training data. T adds the 9,856 RealCommTrain training recordings while retaining the original digital augmentation. A+T combines the same real-call data with RealComm-Aug on digital samples. Development recordings are kept out of training. We do not additionally insert the matched digital source subset from DigitalPool into these experiments. A+T is initialized from F, rather than obtained by continuing an already adapted A or T model.

Training uses 16-kHz mono waveforms with a 64,600-sample length limit and the original model-specific cropping and padding rules. The digital mixture retains its type-based sampler, with at most 1,000 utterances per type in each epoch. For T and A+T, all 9,856 call recordings enter the epoch sampling pool through 88 condition--class types; they are appended to the digital samples without additional call-data oversampling. The optimizer and learning-rate schedule follow the original model-specific recipe. In the Teffic-Audio A and A+T runs, AdamW uses an initial learning rate of $4\times10^{-6}$, cosine decay, an effective batch size of 128, and 10 epochs; checkpoints are selected using the existing public-data validation procedure. Each reported result corresponds to one checkpoint, not an average over training runs.

The model architecture and model-specific training recipe are maintained within each comparison. Existing model-internal neural-codec augmentation is retained as part of that recipe; EnCodec~\cite{encodec} is not a newly added RealComm-Aug stage. Data-side bypass for real recordings therefore describes only the new waveform pipeline, not a claim that every operation internal to the training model is disabled. No augmentation is applied at test time.

The documented staged recipe retains original waveforms with probability 0.20 and allocates 0.08, 0.06, and 0.06 to acoustic-only, device-output, and file-output scenarios. Direct-call and acoustic-call scenarios each have probability 0.30. Operation parameters are sampled within the chosen scenario, so the route probability is different from the marginal probability of any individual operation. The model-specific checkpoints and their provenance are retained with the experimental results.

\section{Adaptation Results and Analysis}
\label{sec:adapt}
\subsection{Overall Adaptation and Digital-Domain Retention}
\begin{figure*}[t]
\centering
\includegraphics[width=\textwidth,trim=0bp 8bp 0bp 0bp,clip]{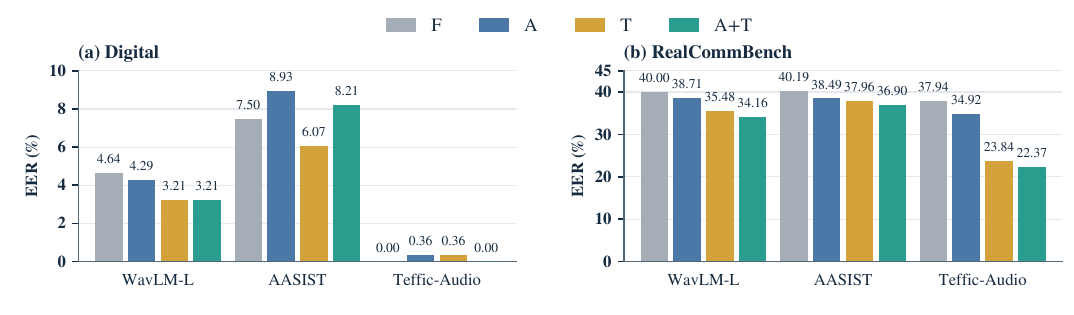}
\caption{Adaptation EER (\%) on Digital/test and RealCommBench (lower is better). Bar labels give two-decimal EERs. The panels use different vertical ranges, both starting at zero. All adapted systems start from F.}
\label{fig:adapt}
\end{figure*}

\textbf{Real-call training provides larger gains than simulation alone.} Figure~\ref{fig:adapt} shows that A lowers call EER by 1.29, 1.69, and 3.02 points for WavLM-L, AASIST, and Teffic-Audio, respectively. T produces larger reductions of 4.52, 2.23, and 14.11 points. Teffic-Audio benefits most from real-call training, moving from 37.94\% to 23.84\%.\footnote{The adaptation comparisons use a matched FP32 re-evaluation of the same Teffic-Audio F checkpoint. Its call EER is 37.94\%, compared with 37.95\% in the frozen zero-shot evaluation shown in Fig.~\ref{fig:zero}(a). Both results are retained with their inference metadata.}

\textbf{Combining simulation and real recordings yields the lowest pooled call EER for every model.} The EERs are 34.16\% for WavLM-L, 36.90\% for AASIST, and 22.37\% for Teffic-Audio. Relative to T, the additional improvements are 1.33, 1.06, and 1.46 points. These consistent overall gains support retaining simulated digital-domain variation when real-call data are available. 

\textbf{Call-domain improvement can come with a digital-domain tradeoff.} WavLM-L has 3.21\% digital EER under both T and A+T, compared with 4.64\% under F. Teffic-Audio remains at or below 0.36\% across all settings. AASIST changes from 6.07\% under T to 8.21\% under A+T. Its call-domain gain therefore comes with a measurable digital-domain tradeoff. Reporting only the call result would conceal this behavior.

\subsection{Input Configuration and Denoising}

\begin{figure*}[t]
\centering
\includegraphics[width=\textwidth,trim=0bp 10bp 0bp 9bp,clip]{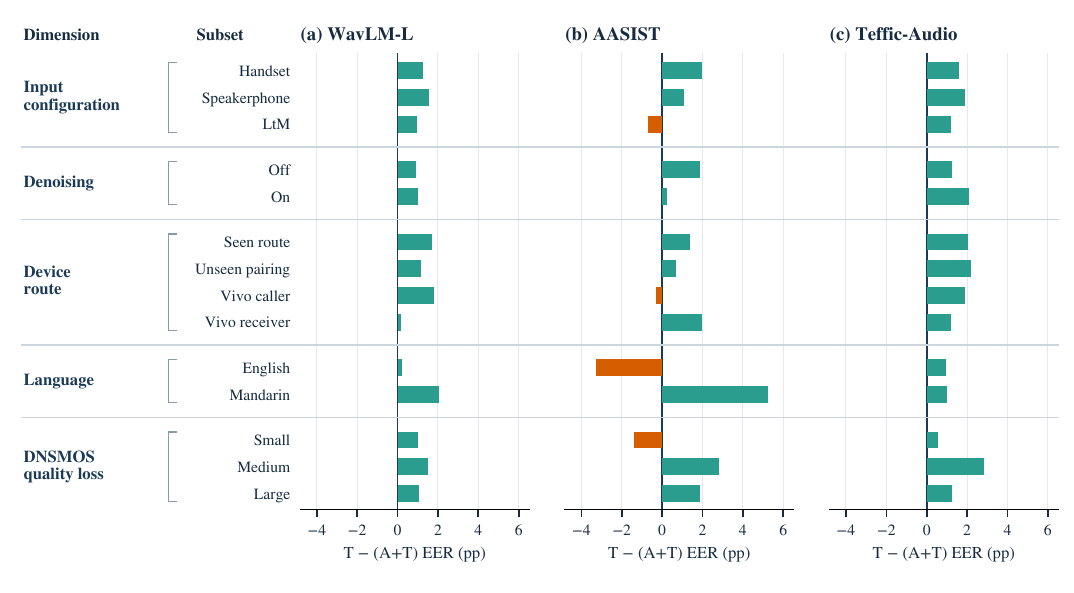}
\caption{Incremental benefit of combined adaptation over real-call training: T minus A+T EER, in percentage points. Positive teal bars denote improvement; negative orange bars denote regression. Labeled blocks group the subsets by input configuration, denoising state, device route, language, and DNSMOS quality-loss stratum. All groups are defined independently of detection scores.}
\label{fig:adaptgains}
\end{figure*}

\textbf{The combined benefit covers all input configurations for WavLM-L and Teffic-Audio.} In Fig.~\ref{fig:adaptgains}, WavLM-L improves from T to A+T by 1.25 points on handset, 1.54 on speakerphone, and 0.98 on LtM. Teffic-Audio also improves in all three configurations. AASIST instead improves on the two OtA subsets while LtM worsens from 23.57\% to 24.29\%. The benefit of combination is therefore broad but not uniform.

\textbf{Acoustic injection remains difficult after adaptation.} For every model and setting, LtM EER is lower than both OtA EERs. Even Teffic-Audio A+T, the strongest overall system in this comparison, reaches 27.86\% on handset and 28.26\% on speakerphone, versus 7.20\% on LtM. The remaining gap indicates that the current combination has not resolved the effects associated with external playback and acoustic capture.

\textbf{The A+T benefit extends to both denoising states for all three models.} Its advantage over T is not confined to a single switch setting.

\textbf{The effect of activating denoising can change after adaptation.} Teffic-Audio F has lower EER with denoising on, whereas T has lower EER with it off. A+T nearly equalizes the two, at 23.63\% and 23.73\%. Robustness to processing states and an average benefit from activating a particular state are distinct properties.

\subsection{Held-Out Device, New Routes, and Languages}
\textbf{Real-call adaptation improves performance on both covered and held-out routes.} T and A+T improve on F in all four route groups for each model. WavLM-L and Teffic-Audio additionally improve from T to A+T in every group. AASIST improves in three groups but changes from 37.54\% to 37.86\% when Vivo is the caller. The results demonstrate transfer beyond the route shared with training, including the held-out Vivo device. Their scope is this particular device ring; they do not establish performance on arbitrary new hardware or communication services.

\textbf{The combined benefit is bilingual for WavLM-L and Teffic-Audio.} Both improve from T to A+T in English and Mandarin. For AASIST, Mandarin EER falls from 42.47\% to 37.23\%, while English rises from 33.20\% to 36.48\%. Thus, although the adaptation benefit is not restricted to one language across the study, AASIST exhibits a language-dependent tradeoff that its pooled result obscures.

\subsection{Quality Strata and Individual Conditions}
We use the quality measurements from Section~\ref{sec:quality} to define small, medium, and large mean DNSMOS-decrease groups, each containing 14 conditions and 7,840 recordings. This score-independent grouping is shared by all models and settings.

\textbf{WavLM-L and Teffic-Audio benefit from A+T at every quality-loss level.} Both improve over T in all three strata. AASIST improves in the medium- and large-decrease groups but worsens in the small-decrease group.

\textbf{Greater quality loss remains associated with higher EER after adaptation.} For all three models under all four settings, EER rises from the small- to the large-decrease stratum. Teffic-Audio A+T reaches 10.97\% in the small-decrease group and 31.20\% in the large-decrease group. This stratification summarizes a mixture of path, route, and denoising effects, rather than identifying an independent causal effect of perceived quality.

\textbf{Pooled improvement coexists with regressions in individual conditions.} Across the 42 recording conditions, A+T improves on T in 25 conditions for WavLM-L, 27 for AASIST, and 28 for Teffic-Audio. It worsens in 12, 11, and 6 conditions, respectively; the remaining conditions tie. The best pooled EER therefore coexists with local regressions, including for the two pretrained-encoder systems.

\subsection{Performance on Public Test Sets}
We additionally evaluate F, A, T, and A+T on ASVspoof 2021 LA~\cite{asvspoof21}, ASVspoof 5~\cite{asvspoof5}, and In-the-Wild~\cite{inthewild}, computing EER independently for each set. The adapted-score evaluation contains 148,176, 680,774, and 31,779 utterances, respectively; ASVspoof 2021 LA uses its evaluation partition. F baselines are the EERs reported by the model contributors, whereas A/T/A+T results are recomputed from the supplied prediction files.

\begin{figure*}[t]
\centering
\includegraphics[width=\textwidth,trim=0bp 21bp 0bp 5bp,clip]{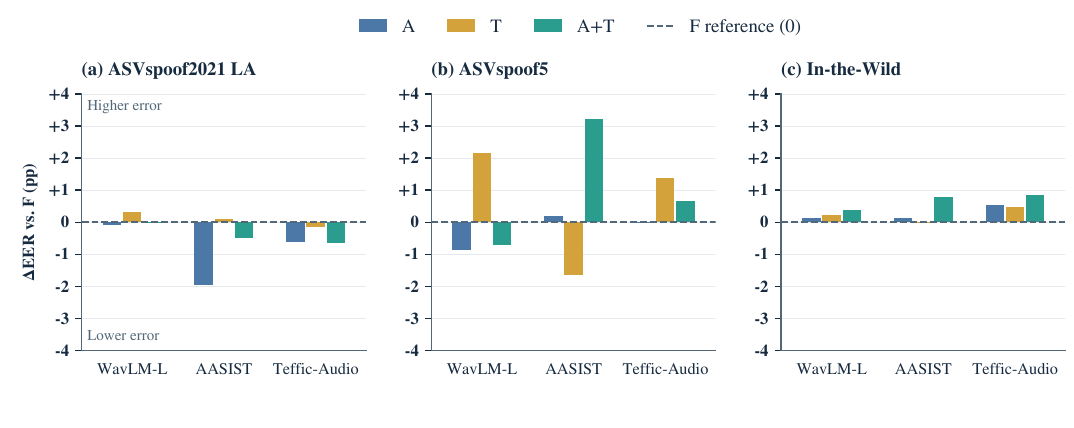}
\caption{Public-test EER changes relative to F, in percentage points. Each bar is the EER of A, T, or A+T minus the same model's F baseline on the same test set. Negative values indicate improvement. The three panels share a $[-4,4]$ vertical range, and the dashed line denotes no change from F.}
\label{fig:publicretention}
\end{figure*}

\textbf{Simulation benefits some external test sets as well as the call domain.} All three models improve on ASVspoof 2021 LA under A. On ASVspoof 5, WavLM-L improves by 0.85 points, while AASIST and Teffic-Audio change by $+0.20$ and less than 0.01 points, respectively. These results indicate useful transfer in some external conditions, rather than a uniform improvement across datasets.

\textbf{Combined adaptation does not preserve performance on every external test set.} Its ASVspoof 2021 LA EER is nearly unchanged for WavLM-L (1.837\% to 1.849\%) and lower for AASIST and Teffic-Audio. However, In-the-Wild EER increases by 0.38, 0.78, and 0.85 points for the three models; AASIST and Teffic-Audio also regress on ASVspoof 5. Target-domain gains should therefore be assessed together with retention on external test sets.

\subsection{Model-Dependent Response to Augmentation}
\textbf{WavLM-L and Teffic-Audio respond more consistently across broad subgroups.} In Fig.~\ref{fig:adaptgains}, A is better than or tied with F in every displayed input, denoising, route, language, and quality subgroup, and A+T is better than T. AASIST also improves overall, but some group gains are accompanied by losses elsewhere, including English and LtM under A+T.

\textbf{Capacity and pretraining are possible explanations, not isolated mechanisms.} WavLM-L and Teffic-Audio use large pretrained speech backbones, while AASIST has a compact task-specific architecture. Richer initial representations and greater capacity may help accommodate channel variation while retaining authenticity cues. This interpretation is consistent with prior corruption analyses~\cite{robustness}, but the present comparison does not isolate these factors, since architecture and pretraining change together with capacity. It therefore supports a model-dependent response, not a demonstrated capacity bottleneck.

\section{Discussion and Conclusion}
RealComm exposes a gap between detecting a digital source file and detecting its received mobile-call version. The paired design connects that gap to acquisition conditions without replacing the source content. Across the evaluated detectors, acoustic injection is consistently more difficult than wired injection. Quality deterioration consistently tracks condition-level detection degradation in the evaluated detectors, whereas high-frequency energy changes alone have limited explanatory value. These observations link detection difficulty to measured signal changes without establishing a causal mechanism.

The adaptation experiments show complementary practical benefits. Simulated augmentation improves the three evaluated models; real-call training provides larger gains; combining the two yields the lowest pooled call EER for each model. Gains extend to unseen route combinations and the held-out device in the current design. Remaining errors are concentrated in challenging acoustic conditions, and AASIST illustrates that pooled improvement can coexist with digital-domain or subgroup regressions. The three public test sets further show that call-domain gains do not guarantee preservation of external-domain performance.

The acquisition design covers multiple devices, two languages, and seven generation systems shared across partitions. Denoising comparisons refer to the caller's user-controlled setting.

The results support evaluating digital references and real-call recordings together, retaining condition labels, and combining simulated digital variation with actual call examples. RealComm provides the paired data and analysis structure needed to study these questions at the signal, detector, and adaptation levels.

\bibliographystyle{IEEEtran}
\bibliography{references}
\end{document}